\documentclass[a4paper,12pt]{article} 

\usepackage{graphicx}
\usepackage{txfonts}
\usepackage{subfigure}
\usepackage{epstopdf}
\usepackage[backref]{hyperref}
\usepackage{rotating}
\usepackage{natbib}
\usepackage{authblk}
\usepackage{fullpage}
\usepackage[top=2in, bottom=1.5in, left=0.5in, right=0.5in]{geometry}

\hypersetup{colorlinks=true,citecolor=cyan,urlcolor=cyan,linkcolor=blue}

\newcommand{\degres}{$^\circ\:$}

\newcommand{\specialcell}[2][l]{%
  \begin{tabular}[#1]{@{}l@{}}#2\end{tabular}}

\begin{document}

   \title{The PAC2MAN mission: a new tool to understand and predict solar energetic events}


\author[1]{Jorge Amaya\footnote{Corresponding author}}
\author[11]{Sophie Musset}
\author[2]{Viktor Andersson}
\author[3,4]{Andrea Diercke}
\author[5,6]{Christian H\"oller}
\author[7]{Sergiu Iliev}
\author[8]{Lilla Juh\'asz}
\author[9]{Ren\'e Kiefer}
\author[10]{Riccardo Lasagni}
\author[12]{Sol\`ene Lejosne}
\author[13]{Mohammad Madi}
\author[14]{Mirko Rummelhagen}
\author[15]{Markus Scheucher}
\author[16]{Arianna Sorba}
\author[15]{Stefan Thonhofer}
 
\affil[1] {Center for mathematical Plasma-Astrophysics (CmPA), Mathematics Department, KULeuven, Belgium}
\affil[2] {Swedish Institute of Space Physics, Sweden}
\affil[3] {Leibniz-Institut f\"ur Astrophysik Potsdam (AIP), Germany}
\affil[4] {Institut f\"ur Physik und Astrophysik, Universit\"at Potsdam, Germany}
\affil[5] {Faculty of Mechanical and Industrial Engineering, University of Technology, Austria}
\affil[6] {Department for Space Mechanisms, RUAG Space GmbH, Austria}
\affil[7] {Aeronautical Engineering Department, Imperial College London, United Kingdom }
\affil[8] {Department of Geophysics and Space Research, E\"otv\"os University, Hungary}
\affil[9] {Kiepenheuer-Institut f\"ur Sonnenphysik (KIS), Germany}
\affil[10]{Department of Aerospace Engineering, University of Bologna, Italy }
\affil[11]{LESIA, Observatoire de Paris, CNRS, UPMC, France }
\affil[12]{British Antarctic Survey, Natural Environment Research Council, England, UK }
\affil[13]{Micos Engineering GmbH, Switzerland}
\affil[14]{Berner \& Mattner Systemtechnik, Germany}
\affil[15]{Physics Department, University of Graz, Austria }
\affil[16]{Blackett Laboratory, Imperial College London, United Kingdom }

\date{December 2, 2014\footnote{Accepted for publication in the Journal of Space Weather and Space Climate}}


   \maketitle
\abstract{
  An accurate forecast of flare and CME initiation requires precise measurements of the magnetic energy build up and release in the active regions of the solar atmosphere. We designed a new space weather mission that performs such measurements using new optical instruments based on the Hanle and Zeeman effects. The mission consists of two satellites, one orbiting the L1 Lagrangian point (Spacecraft Earth, SCE) and the second in heliocentric orbit at 1AU trailing the Earth by 80$^\circ$ (Spacecraft 80, SC80). Optical instruments measure the vector magnetic field in multiple layers of the solar atmosphere. The orbits of the spacecraft allow for a continuous imaging of nearly 73\% of the total solar surface. In-situ plasma instruments detect solar wind conditions at 1AU and ahead of our planet. Earth directed CMEs can be tracked using the stereoscopic view of the spacecraft and the strategic placement of the SC80 satellite. Forecasting of geoeffective space weather events is possible thanks to an accurate surveillance of the magnetic energy build up in the Sun, an optical tracking through the interplanetary space, and in-situ measurements of the near-Earth environment.
}{}{}{}{}
 


\section{Introduction}

The PAC2MAN mission is the result of an intensive two-week academic exercise performed in the framework of the ``ESA Alpbach Summer School 2013 - Space Weather: Science, Missions and Systems''. The main goal of our project is to understand the origin of energetic solar activity and their impact on the Earth environment. A precise forecast of the initiation of solar flares and coronal mass ejections (CME) and their impact on the Earth environment is still impossible due to the shortcomings of our current scientific knowledge and the inadequacy of the space missions in service today. We propose a new space mission that seeks to fill the gaps in our scientific understanding of space weather. The mission measures the three dimensional structure of the magnetic field in the solar atmosphere, the propagation speed of interplanetary CME (ICME) between the Sun and the Earth, and the plasma properties at 1AU and in the near-Earth environment. The Photospheric And Chromospheric and Coronal Magnetic field ANalyzer (PAC2MAN) is a space mission dedicated to the study of eruptive solar events by analyzing the evolution of the magnetic field in these three layers of the solar atmosphere. The mission follows space weather events from their formation in the Sun to their impact on the Earth environment. All the technical and scientific solutions presented in this paper were designed and evaluated during the two week duration of the school.

It is impossible to forecast with precision the effects of space weather events on the Earth environment using the currently available scientific and engineering tools. The reason some of these events have a stronger effect than others is not completely identified. The impact level of the events is measured by their geoeffectiveness, i.e. their ability to cause geomagnetic storms. These can be measured by different geomagnetic indices like the {\em Dst} or {\em Ap} indices.

Statistical analysis of observations have shown that 70\% of Earth oriented ICMEs cause geomagnetic storms. This percentage gradually reduces to 60\% for limb ICMEs and 0\% for back oriented CMEs \citep[see][]{Gopalswamy:2007}. Studies of the correlation between in-situ measurements and solar disk observations have shown that the majority of the Earth oriented ICMEs originate within the 50\degres central meridian of the Sun \citep[see][]{Cane:2003}. However, their geoeffectiveness can not be forecasted: roughly the same number of events produce quiet ({\em Dst}=-1 nT), weak ({\em Dst}=-30 nT), strong ({\em Dst}=-50 nT) and intense ({\em Dst}=-100 nT) geomagnetic storms \citep[see][]{Mustajab:2013}.

Stronger magnetic storms have been associated with large north-south components of the interplanetary magnetic field (IMF) and high solar wind speeds \citep[see][]{Mustajab:2013}. However, by the time that such parameters can be measured by satellites stationed at the L1 Lagrangian point, the forecast time of how geoeffective they will be is limited to less than 1 hour. The situation is worse if we consider the effects of solar flare radiation traveling from the Sun to the Earth in minutes at speeds close to the speed of light \citep[see][]{Curto:2009}.

To forecast space weather conditions that are hazardous to humans and technology, we need a tool to study the processes of CME initiation and acceleration in the Sun and the high energy particle acceleration and propagation towards the Earth \citep[see][]{Feynman:2000}. It has been observed that magnetic energy build-up plays a major role in the initiation of the events \citep[see][]{Feynman:1995,Reinard:2010}. To understand this initiation process, we must track the evolution of magnetic structures in active regions of the solar photosphere, chromosphere and corona \citep[see][]{Owens:2006}. Observations in the days previous to the events will be used to forecast the exact day of eruption. Accurate measurements of the magnetic topology and energetic conditions in the Sun will be correlated to the in-situ measurements of plasma conditions near the Earth. Improved forecasting models of ICME and flare geoeffectiveness can be derived from these observations.

A detailed description of the scientific background of the mission is presented in section \ref{sec:sciback}. The mission profile and the engineering solutions are presented in section \ref{sec:mission}. Finally, a detailed description of the optical and in-situ instruments is given in section \ref{sec:payload}.

\section{Mission objectives}
\label{sec:objectives}

The proposed mission has the following operational and scientific objectives:
\begin{description}
  \item[\bf Primary objective:] Understand and predict the initiation and development of potentially hazardous CMEs and flares.
  \item[\bf Secondary objective:] Determine the speed and direction of CMEs in order to forecast near real-time solar wind conditions close to Earth.
\end{description}

A successful mission will be characterized by the following accomplishments:
\begin{itemize}
  \item The new measurements of the magnetic field will lead to improved models of CME initiation and propagation \citep[see][]{Luhmann:1998}.
  \item The mission will detect how likely a flare or a CME will emerge in a given region of the Sun, with high statistical significance, 2-3 days before their occurrence, improving the forecast accuracy reported by \citet{Reinard:2010}.
  \item Improved models will correlate solar magnetic conditions with their effects on the near-Earth environment, leading to better forecasts of CME and flare geoeffectiveness.
\end{itemize}

\section{Scientific background}
\label{sec:sciback}


The amount of energy released during the most intense solar events can reach 10$^{32}$ erg, which makes them the most powerful events in the solar system \citep[see][]{Woods:2006,kret:2011,Schrijver:2012,Cliver:2013,Aulanier:2013}. This magnetic energy is stored in non-potential magnetic fields associated with electric currents in the solar atmosphere. The magnetic energy available in an eruptive event is called ``free energy'' and is by definition the difference between the potential and the non-potential energy. Magnetic flux emergence can increase the amount of free energy available in an active region, raising the complexity of the magnetic field and its non-potentiality.

Under such circumstances, an instability can lead to magnetic reconnection and to a topology closer to the potential configuration. During the reconnection process magnetic energy is released that can trigger flares and/or CMEs. The magnetic energy is transferred in the form of particle acceleration, plasma motion (ejection of material) and plasma heating \citep[see][]{Emslie:2004}. To calculate the energy budget of an eruptive event it is necessary to measure the amount of free energy available in the active region before the event and the amount of energy effectively released during the event.

The evolution of the coronal magnetic field and electric currents in active regions is a key observable to predict eruptive events such as flares and CMEs days before their occurrence. The magnetic topology of the Sun is usually measured in the photosphere using the Zeeman effect: the spectral lines observed are split in three components of different polarization states due to the presence of the magnetic field. However, in the corona the Zeeman splitting is usually small compared to the thermal broadening.

One of the most promising methods to measure the coronal magnetic field is the interpretation of the Hanle effect in spectral lines. In the corona, the light of spectral lines formed in the lower layers is scattered, introducing a linear polarization in the 90$^\circ$ direction which can be observed in off-limb coronal structures. The direction and degree of the polarization are modified in presence of a local magnetic field. These modifications depend both on the direction and the strength of the local magnetic field. In addition the Hanle effect can be measured by integration over the whole line profile. This property can be used to observe faint spectral lines. However, the analysis of the Hanle effect in a single line cannot give complete information about the local magnetic field because only two parameters of the polarization are measured. Therefore, Hanle measurements have to be combined with other observations in order to calculate the three components of the magnetic field \citep[see][]{Bommier:1982}. The Hanle sensitivity to the magnetic field strength is not the same for all lines: some lines are useful to determine strong magnetic fields, whereas other lines are adapted to probe weaker fields \citep[see][]{Sahal:1981,Judge:2001}.

\subsection{Previous work}

Today only the photospheric magnetic field is continuously measured whereas the coronal field is deduced from numerical models using the former as boundary conditions. Potential and force-free models are used for the extrapolations \citep[see][]{Cheung:2012}, but this method gives uncertain results due to electric currents and non-force-free conditions in the chromosphere and transition region, as well as small scale currents missing in the numerical extrapolations. The extrapolations are also sensitive to the errors in the boundary conditions (for example, the observational errors in the measurements of the photospheric magnetic field). Reviews of the non-linear force-free extrapolation methods and discussion of their limitations can be found in \citet{Demoulin:1997,Amari:1997,McClymont:1997}

A recent study by \citet{Sun:2012} failed to quantify the magnetic free energy loss during a major flare using this technique. This example shows the limitations of coronal magnetic field extrapolations. The authors of the publication were able to study the evolution of the free energy in an active region before and after the flare. Before the X2.2 flare a progressive increase of the free energy is observed and a sudden decease follows after the occurrence of the flare. The decrease of free energy is equal to (3.4 $\pm$ 0.4)$\times 10^{31}$ erg, which is of the same order of magnitude of the energy observed in accelerated electrons ($5 \times 10^{31}$ erg). This is intriguing since other components of the event must also be powered including the CME kinetic energy, which is usually greater than the non-thermal energy of accelerated electrons \citep[see][]{Emslie:2004}.

\subsubsection{Measurements of the photospheric magnetic field}

The two main instruments recording magnetograms in the photosphere are the Helioseismic Magnetic Imager (HMI) onboard the Solar Dynamic Observatory (SDO) and the Spectropolarimeter in the Solar Optical Telescope (SOT) on Hinode \citep[see][]{Scherrer:2012}. They provide spectropolarimetric measurements in iron lines and their characteristics are listed in Table~\ref{HMI_hinode}. 

These spectropolarimetric measurements lead to a calculation of the vector magnetic field in each spectral line (i.e. at a given altitude), via the interpretation of the Zeeman effect for each line. The magnetic field strength and direction can be calculated from the measurement of the full Stokes parameters in one line profile.

The main difference between the two instruments is that Hinode/SOT scans an active region and thus provides two magnetograms (because two lines are observed) in 45 minutes (duration of scan) for one active region. Because these two lines are close together in wavelength, the magnetograms represent the magnetic field at two close altitudes in the solar photosphere. Therefore Hinode provides the spatial variation of the magnetic field not only in the plane parallel to the solar surface, but also in the direction perpendicular to it. This provides the possibility of calculating the three components of the electric current density from the curl of B. 

SDO/HMI measures the vector magnetic field in only one plane. Gradients of the field can not be obtained in the perpendicular direction to the plane, and therefore only the vertical component of the curl of B can be deduced. However, the time resolution used by the HMI instrument is high enough to follow the evolution of the magnetic structures in active regions. It also makes full-disk magnetograms so all active regions can be observed at the same time.

\begin{table*}[ht]
\centering
\caption{The two main instruments currently providing magnetograms in the photosphere}
\begin{tabular}{lp{0.4\textwidth}p{0.4\textwidth}}

\hline\hline
Instrument     & SDO/HMI & Hinode/SOT/SP \\
\hline
Method         & Images in 6 wavelengths and 4 polarization states & Spatial and spectral scan of region \\
Spectral lines & Fe I at 617.33 nm & Fe I doublet at 630.15 nm and 630.25 nm \\
Field of view  & Full disk         & Active region \\
Spatial resolution & 0.5 arcsec & 0.32 arcsec \\
Time resolution    & 12 min & $\approx$45 min (scan duration) \\
\hline
\end{tabular}
\label{HMI_hinode}
\end{table*}
 
\subsubsection{Measurements of the chromospheric magnetic field}

Chromospheric magnetic fields have been successfully measured using the spectropolarimetric observations in the sodium D1/D2 doublet (at 589.59 nm and 589.00 nm) with the THEMIS Solar Telescope, a ground-based instrument producing spectropolarimetric observations of active regions. The Zeeman effect has been interpreted to produce magnetic field vector maps in the two lines and electric current density vector maps for several active regions \citep[see][]{Bommier:2014}. The use of a doublet is advantageous because it provides the three components of the current density, and not simply the vertical one.

\subsubsection{Measurements of the coronal magnetic field}

Measurements of the coronal magnetic field have been performed along sight lines to radio sources that allow Faraday rotation measurements \citep[see][]{Patzold:1987,Mancuso:1999}. This technique provides information only along the line of sight between the instrument and the radio source, and only by interpolation and careful modeling are able to be used to produce 2D maps of the coronal magnetic field. Other types of measurements have been performed using the Faraday rotation of polarized solar radiation \citep[see][]{Alissandrakis:1995}.

The measurement of magnetic field strength in active regions in the low corona can be inferred with the observation of radio gyrosynchrotron emission \citep[e.g.][]{Gary:1994}. This technique has so far been used only for active regions with strong magnetic field: the measurements are difficult to analyze because the height of the radio sources are difficult to interpret.

The most promising technique that remains is the measurement of the full Stokes parameters (I, V, U, Q) in coronal spectral lines from the interpretation of different effects (longitudinal Zeeman effect, resonance polarization, Hanle effect). These measurements can only be obtained in the corona above the solar limb with a coronagraph \citep[see][]{Judge:2001}.

One of the most promising spectral lines for Hanle effect measurements is the hydrogen Lyman $\alpha$ line. \citet{Bommier:1982} explored the theoretical potential of measurements of the coronal magnetic field using the interpretation of the Hanle effect from Lyman $\alpha$. They concluded that this spectral line was a very promising way to provide Hanle effect analyses but that interpretations needed to be complemented by additional measurements in order to determine the three components of the magnetic field. One interesting option is to additionally measure the linear polarization in forbidden emission lines in the infrared range. 

\citet{Raouafi:2009} confirmed that Hanle effect measurements in UV spectral lines is a promising way of measuring the coronal magnetic field. They argue that the hydrogen Lyman $\alpha$ and $\beta$ lines (121.516 nm and 102.572 nm) could be used to differentiate field strengths and make them complementary to each other.

One of the most promising infrared lines to measure the coronal magnetic field has already been used \citep[see][]{Lin:2004}. Spectropolarimetric measurements of the off-limb corona, in the emission line Fe XIII (1075 nm), have been accomplished with an optical fiber-bundle imaging spectropolarimeter installed in a ground-based solar coronagraph (SOLARC). This forbidden coronal line has a high potential to determine the physical conditions of the plasma with a temperature of about 2 MK. This line is very sensitive to electron density \citep[see][]{Chevalier:1969,Flower:1973}. \citet{Lin:2004} measured the full Stokes linear and circular polarized intensity in this line and produced the first two-dimensional coronal magnetic field map (coronal magnetogram).

Recently, the off-limb magnetic field has been measured with a ground-based instrument, the Coronal Multi-channel Polarimeter (COMP), integrated into the Coronal One Shot coronagraph at Sacramento Peak Observatory \citep[see][]{Tomczyk:2008}. The spectropolarimetric measurements in the forbidden Fe XIII emission lines at 1074.7 nm and 1079.8 nm and the chromospheric line HeI at 1083.0 nm have been used, with the interpretation of the Zeeman and Hanle effects combined in the corona to calculate the vector magnetic field, and to produce 2D maps of the magnetic field.

\subsection{Scientific requirements for the PAC2MAN mission}
\label{sec:scireq}

In order to improve our analysis of the free energy build up and release in the corona, as well as the development of instabilities in the active regions, we need to:

\begin{itemize}
\item improve the extrapolation: include non-linear effects introduced by current systems in the corona using alternative models others than force-free field. This advanced modeling is still difficult to achieve \citep[see][]{Judge:2001}
\item measure continuously the magnetic field in other layers of the sun (chromosphere, transition region, corona) to add more constraints on the magnetic field, current systems and magnetic energy. Perform simultaneously direct measurements of magnetic field vector and electric current density (via the curl of magnetic field) in several layers of the solar atmosphere.
\end{itemize}

Our mission fulfills the second requirement, by providing simultaneous measurements of the magnetic field in the photosphere, chromosphere and low-corona, with a high time cadence and spatial resolution. It provides data to follow the evolution of magnetic field and magnetic free energy in the corona, and then to understand the magnetic thresholds leading to eruptive events. A continuous data recording and a constant communications link with ground stations provide the means to predict the intensity of eruptive events such as flares and CMEs days before their occurrence.

The measurements in the photosphere and in the chromosphere are performed with spectropolarimetric observations in two doublets, to provide the three components of the magnetic field and of the electric current density in these layers. These measurements use the interpretation of the Zeeman effect near the disk center where the projection effects are negligible. The measurements of the coronal field are obtained from the off-limb corona using a coronagraph interpreting the Hanle effect. A detailed three dimensional reconstruction of the photospheric and the chromospheric magnetic fields is possible using a stereoscopic view. To attain this goal, measurements of the center of the solar disk obtained from an L1 orbit can be complemented with observations of the off-limb coronal magnetic field performed from a second spacecraft at nearly 90$^{\circ}$.

The magnetic field reconstructions are compared against images obtained with an UV imager placed in each spacecraft. This instrument provides images of the chromospheric and coronal plasma loops that follow the 3D coronal magnetic field. White light observations (scattered light) are also used to estimate physical properties and the global topology of the corona.

To obtain a complete picture of the full space weather system, additional measurements must be performed in the space between the Sun and the Earth. Using the stereoscopic placement of the satellites to advantage, it is possible to track Earth-directed CMEs. Observations of the scattered light of the ejections with an additional coronagraph and an heliospheric imager allow determination of the exact direction of propagation and velocity of ICMEs. In-situ measurements of the solar wind conditions at 1 AU  provide an additional characterization of the effects of solar events in the interplanetary environment.
\section{Mission profile}
\label{sec:mission}

The PAC2MAN mission is composed of two satellites and a network of ground tracking stations that monitors the Sun and interplanetary space between Sun and Earth from two nearly orthogonal positions: the Spacecraft Earth (SCE) satellite orbits around L1 and the Spacecraft 80 (SC80) is placed in heliocentric orbit trailing the Earth by 80$^{\circ}$.

\subsection{Orbit design}

Among the constraints that guided the placement of the two spacecraft we defined:

\begin{itemize}
\item {\bf Scientific constraints:} a constant stereoscopic view of the interplanetary space from the Sun to the Earth, off-limb measurements of the solar atmosphere, and in-situ measurements of the solar wind at 1 AU and near the Earth.
\item {\bf Engineering constraints:} data transmission rates from the spacecraft, thermal control, power supply, injection orbits, and cost of the mission.
\end{itemize}

\begin{figure*}[h]
\centering
\includegraphics[width=0.4\textwidth]{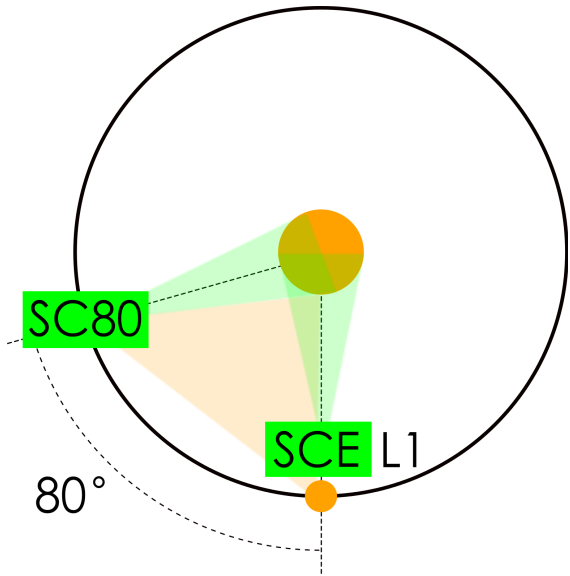}
\caption{Sketch of the orbits of the two spacecraft in the ecliptic plane relative to the Earth and the fields of view of the optical instruments.}
\label{figOrbit}
\end{figure*}

During the design phase different orbits were considered. These considerations led to the following final selected orbits (see Figure~\ref{figOrbit}).

\subsubsection{SC80}

The SC80 follows the same elliptical orbit as the Earth but with a smaller True Anomaly (80$^\circ$ less than the Earth's), i.e. it trails the Earth during its orbit.

To put the spacecraft in orbit two main maneuvers need to be performed. The first one, at Earth's apogee, using a $\Delta$V value of $1.02$ km$\cdot$s$^{-1}$ inserts the spacecraft into an elliptical orbit with an eccentricity $\epsilon  = 0.0678$. The spacecraft makes two lapses in this transfer orbit in order to build up the required tilt. This operation requires a total time of 26.7 months.

The second maneuver applies a $\Delta$V of $1.02$ km$\cdot$s$^{-1}$ in the opposite direction again at the apogee. This second maneuver places the spacecraft in its final orbit and is provided by the spacecraft thrusters.

\subsubsection{SCE}

The second satellite follows a Halo Lissajous orbit around L1 (see parameters in Table~\ref{tableOrbit}). A Soyuz launcher provides the necessary c3 energy to enter the orbit, while 10 N thrusters provide the necessary $\Delta$V for orbit maintenance in this unstable orbit. The estimated transfer time is 3.5 months.

Using STK\footnote{Systems Tool Kit (STK) is a software package commonly used for the calculation of complex orbital dynamics.} we constructed a free non escape orbit through velocity increments along the escape direction showing that a $\Delta$V $\approx$ 13.5 m$\cdot$s$^{-1}$ is required for orbit maintenance. Comparing those values to the ones of SOHO and Herschel of 2.4 m$\cdot$s$^{-1}$ and 1 m$\cdot$s$^{-1}$ respectively, a nominal value of $\Delta$V $\approx$ 10 m$\cdot$s$^{-1}$ per year was assumed for orbit maintenance. For the nominal mission lifetime of 6 years this translates to a $\Delta$V $\approx 60$ m$\cdot$s$^{-1}$. Corrections during orbit entrance require an additional $\Delta$V $\approx 65$ m$\cdot$s$^{-1}$, adding to a total $\Delta$V $\approx 125$ m$\cdot$s$^{-1}$.

\begin{table*}[h]
\centering
\caption{Parameters associated to the unstable Halo Lissajous orbit of SCE around the L1 point.}
\begin{tabular}{ccc}
\hline\hline
Axis & Amplitude of oscillation (km) & Orbital period (days) \\
\hline
X & $2.6*10^{5}$ & $177.566$ \\
Y & $8.3*10^{5}$ & $177.566$ \\
Z & $4*10^{5}$ & 184 \\
\hline
\end{tabular}
\label{tableOrbit}
\end{table*}

\subsection{Spacecraft Design}
\begin{figure*}[h]
\centering
\includegraphics[width=0.8\textwidth]{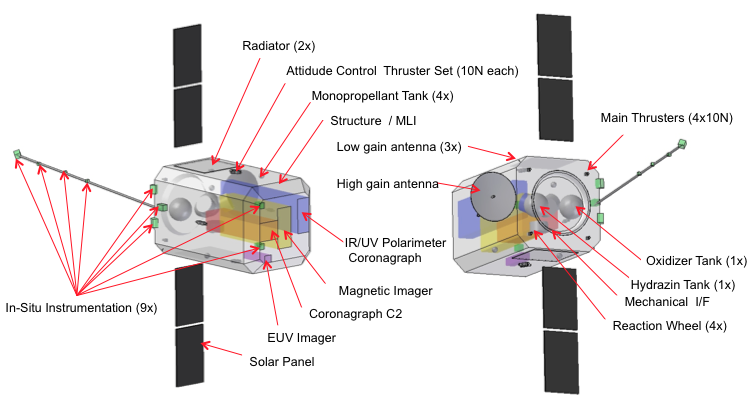}
\caption{Two views of the SCE. Some of the plasma instruments are placed in a 6 m boom behind the spacecraft, shadowed from the Sun. Optical instruments and low energy solar wind detectors point towards the Sun.}
\label{figSatelliteSCE}
\end{figure*}

\begin{figure*}[h]
\centering
\includegraphics[width=0.8\textwidth]{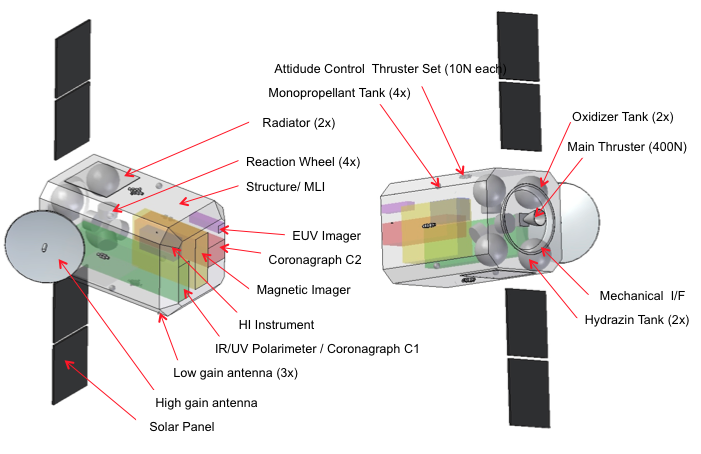}
\caption{Two views of the SC80. Optical instruments point towards the Sun. A main thruster is used to inject the satellite in the final orbit. Only one in-situ instrument is installed onboard this satellite facing the Sun.}
\label{figSatelliteSC80}
\end{figure*}

3D-CAD models of the two spacecraft are shown in Figures~\ref{figSatelliteSCE} and~\ref{figSatelliteSC80}. The SC80 presents a more complex design required for the telemetry, tracking, and command (TT\&C) and propulsion subsystems, but the remaining subsystems are very similar.

A solar wind analyzer and optical instruments on SC80 point towards the Sun with a narrow field of view. The only instrument pointing in a different direction is the HI Instrument which has a field of view of 42$^\circ$ with a pointing to 29$^\circ$ elongation. The SCE in-situ instruments are placed following the measuring constraints, which include magnetic cleanliness, orientation in the parker spiral direction and shadowing from the Sun.

\begin{figure*}[h]
\centering
\includegraphics[width=0.6\textwidth]{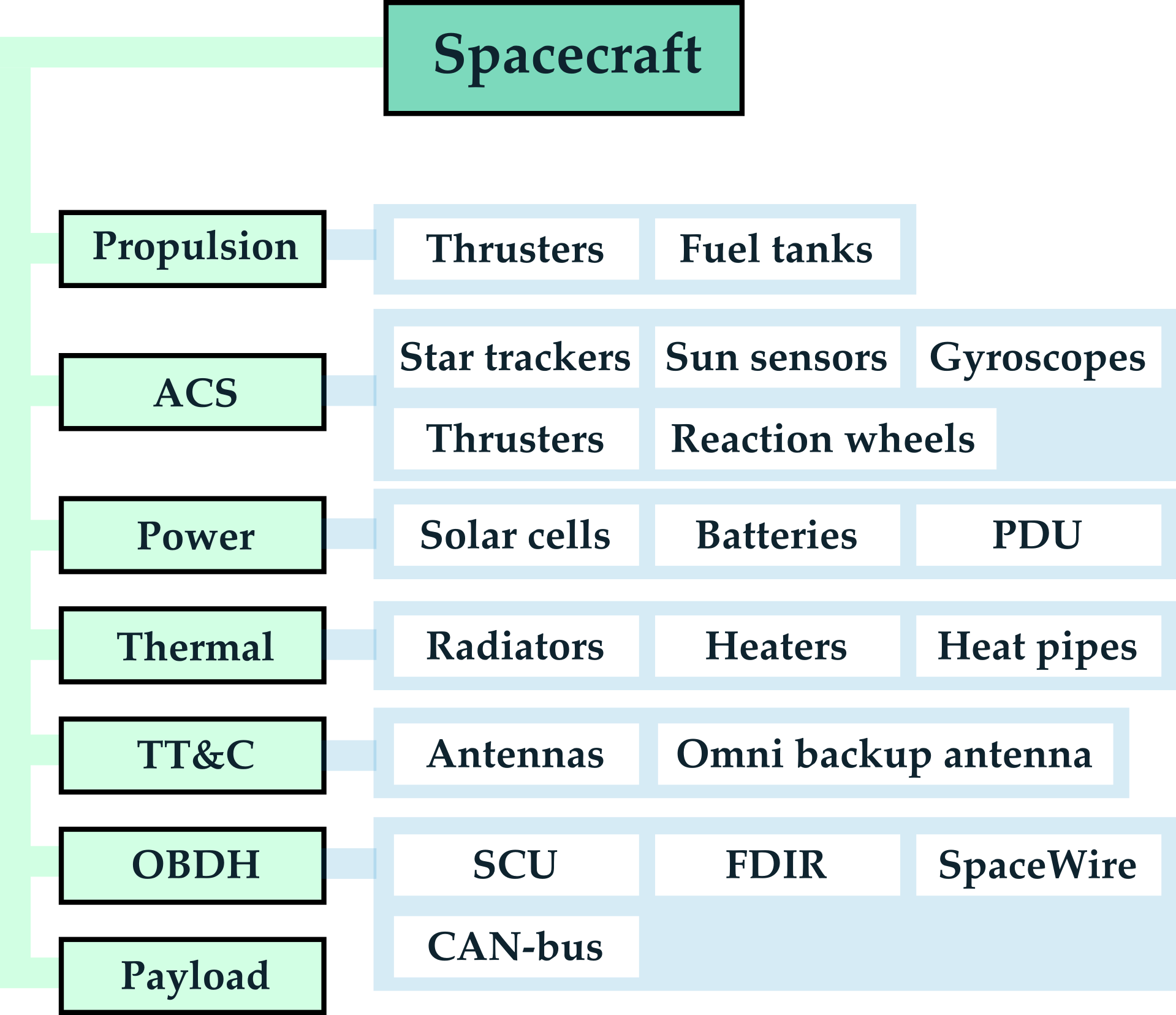}
\caption{Block diagram of the different spacecraft subsystems.}
\label{figDesign}
\end{figure*}

Figure~\ref{figDesign} shows the main components of the spacecraft. The subsystems were analyzed in detail and in the following sections we present the selected characteristics for each of them. Notice in particular the difference in the sizes of the antenna and the propellant tanks, and the inclusion of a boom in the SCE.

\subsubsection{Propulsion}

Engine design is dictated by the $\Delta$V necessary for orbit acquisition and represents a fine balance between fuel efficiency, thrust requirements, mass and reliable operational time. The mission requires one primary engine for the SC80 orbit injection and several secondary engines, on both spacecraft, for attitude control and orbit corrections. Several possible propulsion systems were considered, including those with a low Technology Readiness Level (TRL) like electric propulsion, but we chose to use a standard chemical bi-propellant. All engines use monomethylhydrazine (MMH) as fuel, oxidized with dinitrogen tetroxide ($N_2$$O_4$). Both chemicals are stored in tanks of different volume (Astrium OST 01/X for SC80 and OST 31/0 for SCE). The volume ratio between the fuel and oxidizer is $1.65$. The system consists of eight pairs of Astrium S10-21 thrusters ($12$ + $4$) for redundancy. The SCE is outfitted with 4 additional 10 N thrusters, while SC80 incorporates a single primary Astrium Apogee S$400-15$.

\subsubsection{ACS}

The attitude control system (ACS) is used to perform attitude corrections to maintain the pointing accuracy within the required limits of the optical instruments. The sensors consist of a High Accuracy Star Tracker system (HAST) composed by two Star Sensor Heads (SSH) and one Star Sensor Electronics Unit (SSEU), an Inertial Reference Unit (IRU) and two Sun Sensors (SS). The actuators consist of 4 Reaction Wheels (RW) and a set of 16 Hydrazine Thrusters.

The ACS was designed to fulfill the requirements of the most sensitive optical instrument, the Magnetic Imager, that requires an accuracy of $0.5$ arcsec and an exposure time of $4$ s during science mode. These requirements can be met by combining the HAST and IRU systems, with the first one giving a $0.1$ arcsec pointing accuracy before exposure and the second giving the rate errors to interpret the data with a bias drift of $0.0005$ arcsec per second. This allows for a total drift of $0.1$ arcsec during exposure time. The main ACS control modes are the following:

\begin{itemize}
\item \textbf{Coarse Sun Acquisition:} used right after the injection into orbit and during orbit maintenance operations. The SS provides the attitude and the RW slew the vehicle to the Sun. In case of saturation of the wheels, thrusters are used to offload the stored momentum.
\item \textbf{Fine Sun Acquisition:} the star tracker in the HAST system provides information to the RW in order to put boresight within $0.5$ arcsec from the target.
\item \textbf{Science Mode:} star trackers provide pitch and yaw errors while the IRU provides rate errors for data handling. The RW maintains the required pointing and stores the daily momentum build up from environmental torques.
\item \textbf{Safe Mode:} only the SS and the IRU provide attitude determination to maintain a low power consumption and ensuring a coarse pointing for thermal safekeeping.
\end{itemize}

The total torque applied by external perturbations was calculated to estimate the ACS fuel consumption. In the selected orbits only two perturbations are important: magnetic torque and Solar Radiation pressure torque. Assuming the presence of a magnetic field of 1 $\mu$T (WCS), a dipolar charging for the spacecraft of 1 A$\cdot$m$^{-2}$, a reflectivity of $0.6$mm and a displacement of $0.5$ m between the center of mass and the center of radiation pressure a value of 31.7 $\mu$Nm was found (with a $50\%$ margin). Taking this torque into account and knowing the moments of inertia we estimated that in order to keep the pointing within the accuracy threshold of $0.5$ arcsec the RW need to be used for adjustments only after 10 s. With an exposure time of 4 s the accuracy is guaranteed. Considering a thruster torque of $7.5$ Nm the amount of fuel necessary for corrections is $1.12$ kg per year (including a $100\%$ margin).

\subsubsection{Power}

The power subsystem provides, stores and distributes the necessary power for the spacecraft bus and payload operations. Solar arrays have been dimensioned with their end of life performance at a distance of 1 AU from the Sun. The power storage system supports each spacecraft in the worst case scenario of control loss for up to 12 hours. This also includes a power distribution unit (PDU). The overall power consumption for both spacecraft has been identified as non-critical for the final design. Table~\ref{tablePower} show that the main power consumer for SC80 is the communications subsystem, while payload is the main power consumer for SCE. The solar panel area is 2.2 m$^{2}$ for SCE and 2.7 m$^{2}$ for SC80.

\begin{table*}[h]
\centering
\caption{Power budget for each spacecraft.}
\begin{tabular}{lcc}
\hline\hline
Subsystem & SCE (W) & SC80 (W) \\
\hline
Payload & 226 & 109 \\
Propulsion & 21 & 24 \\
ACS & 42 & 47 \\
TT$\&$C & 60 & 350 \\
OBDH & 21 & 24 \\
Thermal & 21 & 24 \\
Power & 63 & 71 \\
Total consumption & 436 & 557 \\
With 20$\%$ margin & 523 & 668 \\
Total power available after 11 years & 654 & 823 \\
\hline
\end{tabular}
\label{tablePower}
\end{table*}

\subsubsection{Thermal}

The spacecraft are maintained at a room temperature of 20 $^\circ$C. Some of the remote sensing instruments have a low operating temperature and require active cooling, but these are addressed by the relevant payload design. Both satellites are assumed to have one constantly Sun-facing side. The thermal radiation and the solar reflection from the Earth are considered negligible. Heat pipes provide an effective passive heat transfer from internal heat sources and from the Sun-facing side to the radiators.


\subsubsection{Telemetry, tracking and control}

A link-budget was calculated for each satellite in order to design a telemetry, tracking and control subsystem (TT$\&$C) which fulfills the mission requirements.

Both satellites operate at a frequency of 8500 MHz (X-Band). Even though Ka-Band provides larger antenna gains, X-Band transmissions are less demanding regarding elevation, angle or phase shift compensation of ground stations. Three 15 m X-Band ESTRACK ground stations allow for continuous communications with the spacecraft. Each of them has an approximate relative alignment of 120$^\circ$. Continuous communication reduces the necessary down-link bit rate.

Reliable communications are obtained using a code rate of $1/2$. Even though this doubles the necessary data rate (-3 dB) it leads to an additional coding gain of 6 dB. BPSK modulation is used in order to minimize the bit energy to noise ratio demand to the least possible value (9.6 dB for a maximum bit error rate of $10^{-5}$).

Both satellites use high gain, narrow beam, parabolic antennas for scientific and operational data. In addition low gain omnidirectional antennas are used in case of a failure of the main antenna. The noise at the receiving antenna at the X-Band frequency causes an additional link degradation of $15.2$ dB.


The link-margin, generally required to be above 3 dB, was calculated for both satellites and is shown in Table~\ref{tableTTC}.

\begin{table*}[h]
\centering
\caption{Link budget calculations used for communications analysis.}
\begin{tabular}{lcc}
\hline\hline
Spacecraft & SCE & SC80 \\
\hline
Daily data volume & $43.2$ GB & 910 MB \\
Data rate & $10485$ kbit$\cdot$s$^{-1}$ & $200.6$ kbit$\cdot$s$^{-1}$ \\
Transmission power & 15 W & 160 W \\
Antenna diameter & $0.5$ m & $2.3$ m \\
Distance & $1.5$ Mkm & 193 Mkm \\
Link-margin & $4.6$ dB & $3.2$ dB \\
\hline
\end{tabular}
\label{tableTTC}
\end{table*}

\subsubsection{On-Board Data Handling}

The On-Board Data Handling subsystem (OBDH) consists of dedicated control units for each module. A Satellite Control Unit (SCU) handles the housekeeping, such as regular temperature and pressure measurements, as well as Failure Detection, Isolation and Recovery (FDIR) based on the derived information from other sub-control units. The SCU interprets and executes commands which are forwarded by the communication-control subsystem and handles the data before it is sent to Earth. The data is recorded on a solid state mass memory which can store up to 70 GB.

Two separate modules for the Service (SVM) and the Payload (PLM) report back to the SCU. The PLM consists of a payload controller, a data processing unit and thermal controlling subsystem. The SVM consists of an attitude and orbit controller, a propulsion controlling unit, a communication controller as well as a controlling unit for the thermal system. A dedicated power control unit is used to manage the power source, the power storage and the power distribution. The control units belonging to the PLM are attached to a high speed SpaceWire network while the controllers of the service module use a CAN-Bus to communicate within the satellite.

\subsection{Mass budget}

The mass of each one of the subsystems is based on available information of past missions, published catalogs and personal experience from expert consultants. The mass budget was calculated for each individual subsystem considering safety margins as shown in Table~\ref{tableMass}.

\begin{table*}[h]
\centering
\caption{Mass budget of each spacecraft. Values are given in kg.}
\begin{tabular}{lccc}
\hline\hline
           & Safety margin & SCE & SC80 \\
\hline
Payload & $20 \%$ & $206.7$ & $204$ \\
Propulsion & $5 \%$ & $14.4$ & $110.3$ \\
ACS & $5 \%$ & $90.3$ & $90.3$ \\
TT$\&$C & $5 \%$ & $52.5$ & $88.0$ \\
OBDH & $10 \%$ & $22.0$ & $33.0$ \\
Thermal & $10 \%$ & $42.4$ & $56.2$ \\
Power & $5 \%$ & $62.2$ & $75.0$ \\
Structure & $5 \%$ & $127.2$ & $168.5$ \\
System margin &  $25 \%$ & $154.1$ & $206.3$ \\
\hline
S/C Dry mass &   & $770.9$ & $1031.6$ \\
Propellent &  & $86.1$ & $496$ \\
S/C Wet mass & $5 \%$ & $857.0$ & $1527.6$ \\
Adaptor &  & $150.0$ & $150.0$ \\
\hline
Launched mass &  & $1007.0$ & $1677.6$ \\
Launcher potential &  & $2150.0$ & $2150.0$ \\
Launcher margin &  & $1143.0$ & $422.4$ \\
\hline
\end{tabular}
\label{tableMass}
\end{table*}

\subsection{Launchers}

The mission is designed for an operational lifetime of 6 years covering a period of maximum solar activity. The satellites will reach their final orbit 3 years before the predicted maximum of solar cycle 26. Delays of the launch date could lead to operations during solar minimum bellow optimal operation conditions. Operations in such case will still allow measurement of up to $0.5$ CMEs/day \citep[see][]{Gopalswamy:2003}.

Two Soyuz rockets taking off from Kourou, French Guiana, are used to launch the spacecraft to the required transfer orbits. The first launcher injects SC80 into an elliptic transfer orbit through an Earth escape orbit with c3 =$1.15$ km$^2\cdot$s$^{-2}$. The second launcher injects SCE into its Lissajous orbit through an Earth escape orbit with c3 = $0.08$ km$^2\cdot$s$^{-2}$.

\subsection{Cost estimation}

Because we propose using two spacecraft, the Rough Order of Magnitude (ROM) budget for the full mission is similar to an ESA L-class mission. Many features in the design can drive down the total cost of the mission, including a possible platform heritage from Planck/Herschel, architectural similarities between the spacecraft, size of the satellites and a possible shared launch for the SCE. Descoping options were also evaluated, but using current technology descoping is not recommended. A six year mission extension with a more operational Space Weather focus will add an extra 100 million euros to the cost.

\begin{table*}[h]
\centering
\caption{The ROM cost estimation for the full PAC2MAN mission.}
\begin{tabular}{lc}
\hline\hline
Activity & Cost (M Euro)\\
\hline
Launchers & 120 \\
SCE platform & 250 \\
SC80 platform & 200 \\
SCE operations & 40 \\
SC80 operations & 50 \\
Ground segment & 100 \\
ESA cost & 760 \\
Payload instruments and scientific data processing & 420 \\
Total cost & 1180 \\
Agency/management (included in total cost) & 130 \\
\hline
\end{tabular}
\label{tableCost}
\end{table*}

\begin{figure*}[h]
\centering
\includegraphics[width=0.7\textwidth]{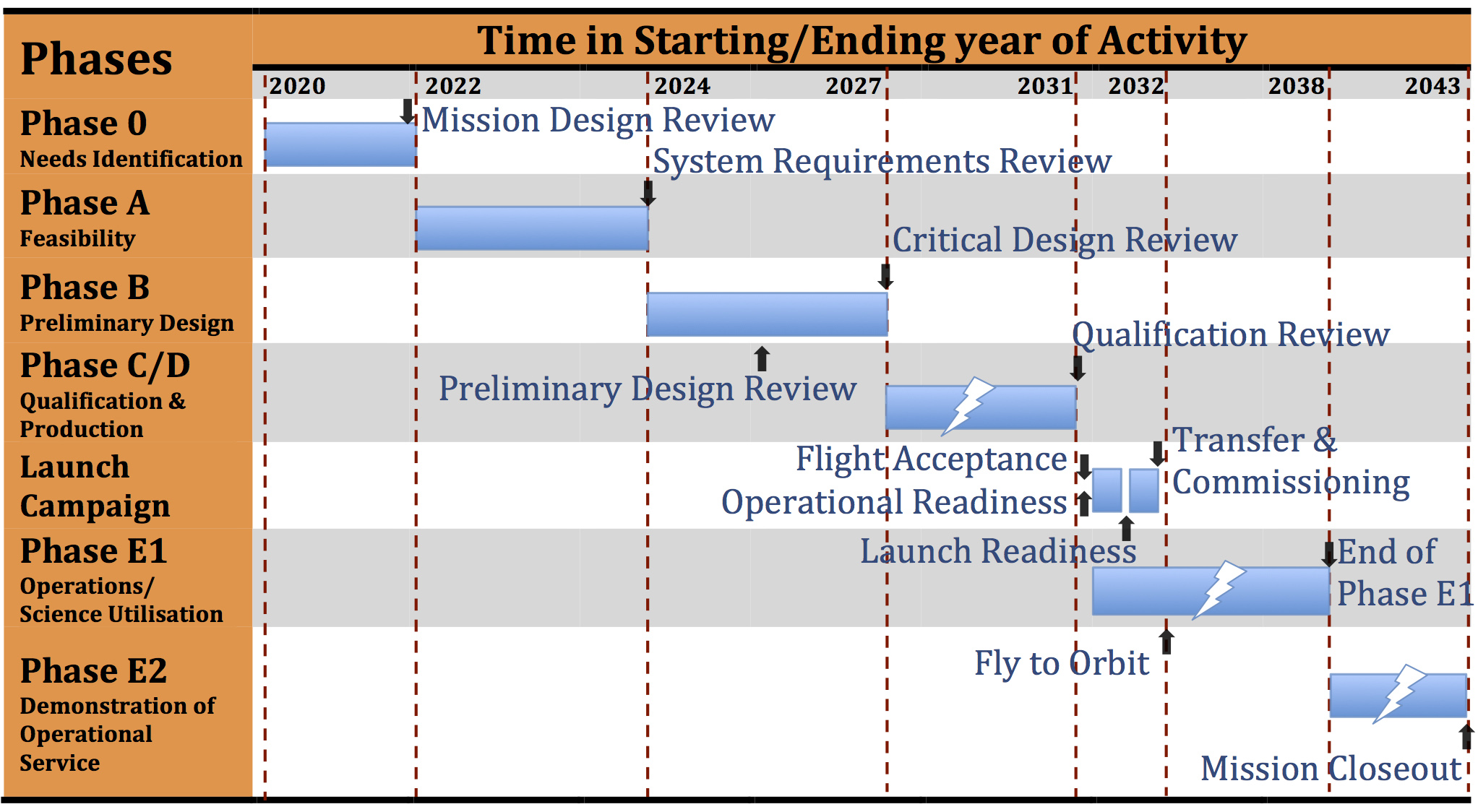}
\caption{Planed schedule for the construction and operation of the PAC2MAN mission.}
\label{figProject_plan}
\end{figure*}

\subsection{Risk assessment}

Several scenarios representing different risks for the mission were evaluated and graded based on their likelihood (from low A, to high E) and their impact (from low 1, to high 5). A1 represent the lowest risk and E5 the highest. The most important risks are:

\begin{itemize}
\item Unavailability of scientific instruments close to launch time (B3). The IR and UV coronagraph (UVIRC) and the Multi-Magnetic Imager (MMI) have a TRL of 3. To mitigate this risk we planned for a close monitoring of the instrument developments and additional options in case of mission delay.
\item Degradation of instruments due to constant solar exposure (B3). Especially a degradation of the Polarimeter's filters. We propose to include in the final design a mechanism to switch spare filters in the filter wheels.
\item Loss of SC80 spacecraft (A4). Although a critical risk, the SCE alone carries observational and in-situ instruments more capable than those on the ACE spacecraft.
\item Loss of attitude control (C1). A customized Safe Mode and redundant sensors and actuators are implemented in both systems. One common issue associated to loss of attitude control is the inability to correctly point the solar panels, reducing the available power and forcing the use of batteries designed to last 12 hours. To ensure attitude control a computationally demanding algorithm must be activated using CPU time available when the non-vital systems are turned off.
\item Loss of SCE spacecraft (A3). In this scenario we will use already existing satellites at L1 to cover some measurements of SCE. SC80 provides Space Weather forecasting by measuring CMEs from Sun to Earth from that viewpoint.
\end{itemize}
\section{Payload}
\label{sec:payload}

\subsection{Ultraviolet and Infrared Coronagraph (UVIRC) for spectropolarimetry in coronal lines}
\label{sec:UVIRC}

To calculate the three components of the magnetic field at different altitudes, for a large range of magnetic strengths, we need spectropolarimetric measurements in several lines. Our two satellites are equipped with two identical coronagraphs for spectropolarimetric measurements in infrared and ultraviolet spectral lines, as well as visible light. We detail here the choice of the lines observed to ensure a good measurement of magnetic fields in the corona, and then the design of the specialized instrument.

\subsubsection{Choice of spectral lines}
\label{sec:lines}

The hydrogen Lyman $\alpha$ and $\beta$ are among the brightest coronal lines in the far ultraviolet spectrum. They have suitable sensitivity to determine the coronal magnetic field using the Hanle effect \citep[see][]{Raouafi:2009}.
The L$\alpha$ radiation comes mainly from the fluorescent scattering from the intense chromospheric underlying source \citep[see][]{Gabriel:1971,Bommier:1982}, and the domain of sensitivity of this line to the Hanle effect lies in the range 12-600 Gauss. Lyman $\beta$ has an important contribution from the electron collision \citep[see][]{Raouafi:2009}. It is sensitive to weaker field (maximum sensitivity around 15 G). \citet{Bommier:1982} suggested that a complete reconstruction of the three components of the magnetic field can be obtained using a combination of two observations: the first one is the interpretation of the Hanle effect in the L$\alpha$ line, the second one is the determination of the direction of the magnetic field from the linear polarization of a forbidden line formed in the same region.

Our spectropolarimeter measures the full Stokes parameters in the two UV lines Lyman $\alpha$ and Lyman $\beta$, but also in three forbidden emission lines in the infrared. The infrared lines are  FeXIII 1074.7 nm, FeXIII 1079.8 nm, and HeI 1083.0 nm. The two iron lines are forbidden coronal emission lines with a high potential to determine the direction of the coronal field \citep[see][]{Judge:2001}. The helium line is a forbidden chromospheric line, and provides an additional and complementary measure to the chromospheric magnetic field imager using the Zeeman effect in the sodium doublet (instrument MMI).

\subsubsection{Instrument design}

\begin{figure*}
\begin{center}
\includegraphics[width=0.8\textwidth]{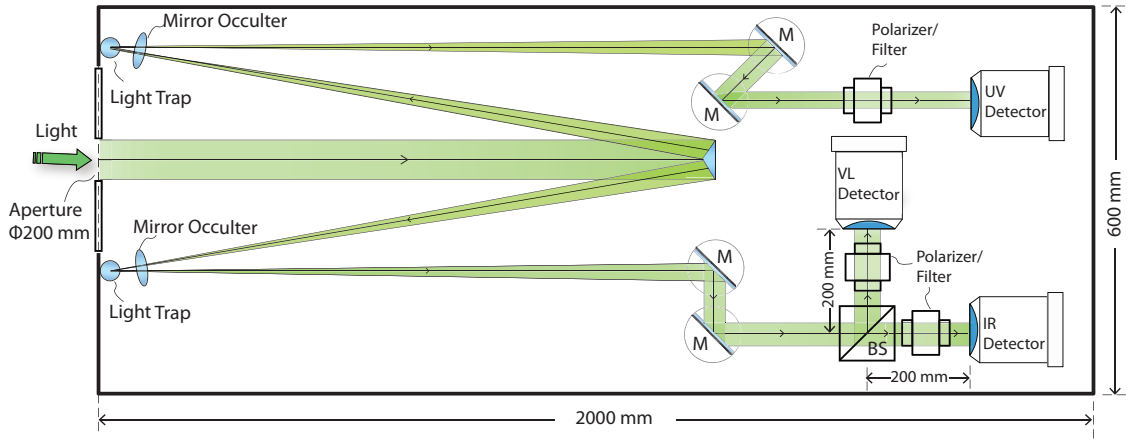}
\end{center}
\caption{Sketch of the spectropolarimeter. Only optical elements needed for the understanding of the principle of the measurement (e.g. internal baffles are not shown).}
\label{spectro}
\end{figure*}

The design of the Ultraviolet and Infrared Coronagraph was inspired by the Visible light and IR Coronagraph (VIRCOR) proposed for the SolmeX mission \citep[see][]{SolmeX} and the MAGnetic Imaging Coronagraph (MAGIC) using a LYOT+ concept \citep[see][]{lyot}.
The sketch illustrating the principles of the instrument is presented in Figure~\ref{spectro}. The image does not contain all components of the instrument (for example, internal baffles needed for straight light reduction are not shown for clarity). The low corona, between 1.1 Rs and 2.0 Rs, is observed with an internally occulted coronagraph with an aperture of 20 cm. The same aperture is used for infrared, visible and ultraviolet light, as it has already been proved to be possible in the Spectral Irradiance Monitor (SIM) instrument on the SORCE satellite \citep[see][]{Harder:2005}. The ultraviolet light is separated from the visible/infrared with a two-side-mirror, the two faces having different coatings. 

The pointing accuracy is constrained by the occultation (1/15 of 1.1 Rs): therefore the instrument requires a pointing accuracy of 72 arcsec. The stability of 3 arcsec has to be kept during the longer exposure time of 20 s.
The infrared lines FeXIII 1074.7 nm, FeXIII 1079.8 nm, and HeI 1083.0 nm are analyzed using a Liquid Crystal Variable Retarder device for both the polarimetry and the tunable wavelength selection, and a narrow-band tunable filter.

A six-stage birefringent filter is used to select different wavelengths. It needs to be maintained at a temperature of 30 $^\circ$C, with a variation of less than 5 mC in 24 hours. The detector is a Teledyne imaging HgCdTe 2048$\times$2048 detector with pixel size of 15 $\mu$m. Observation of the K-corona is possible by diverting the light through a beam-splitter, followed by a tunable broad-band filter and a dichroic linear polarizer. The filtered photons are captured by an APS sensor 2048$\times$2048 with pixel size of 15 $\mu$m. The UV HI Lyman alpha line at 121.6 nm is isolated with a high reflectivity Brewster's angle linear polarizer. An APS sensor 2048$\times$2048 with pixel size of 15 $\mu$m is used to capture the filtered light.


The major issue in coronagraphic observations is the reduction of straight light. This becomes difficult when observing in the ultraviolet range, which is significantly fainter than optical and near-infrared light. Internal baffles as well as photon-traps are used to ensure the straight light reduction. The UV detectors require a special care throughout the mission. It is especially important to ensure the cleanliness of this instrument during the entire mission. The filters are replaced during the mission by a filter wheel mounted at the entrance of the instrument.

 \begin{table*}
    \centering
    \caption[]{Overview of the operational characteristics of the remote sensing instruments.}
    \label{Tab:opt}
    \begin{tabular}{p{0.15\textwidth}p{0.01\textwidth}p{0.01\textwidth}p{0.17\textwidth}p{0.1\textwidth}p{0.1\textwidth}p{0.05\textwidth}p{0.2\textwidth}}
	\hline\hline
Instrument & \rotatebox[origin=c]{90}{Onboard SCE} & \rotatebox[origin=c]{90}{Onboard SC80} & Wavelengths & FOV (R$_{Sun}$) & Detector Size (pix) & Pixel Size (arcsec) & Observable \\
\hline
UV and IR coronagraph & $\times$ & $\times$ & \specialcell{HI Ly$\alpha$ (121.6 nm)\\HI Ly$\beta$ (102.6 nm)\\FeXIII (1074.7 nm)\\FeXIII (1079.8 nm) \\ HeI (1083.0 nm)\\Visible (560 nm)} & 1.1 - 2 & 1024$\times$1024 & 5 & Magnetic Field in low corona \\
\hline
Multi-Magnetic Imager & $\times$ & & \specialcell{FeI (630.15 nm)\\FeI (630.25 nm)\\NaI (589.59 nm)\\NaI (588.99 nm)} & $<$ 1.07 & 4096$\times$4096 & 0.5 & Magnetic Field in photosphere and chromosphere \\
\hline
Magnetic Imager & & $\times$ & \specialcell{FeI (617.3 nm)} & $<$ 1.07 & 4096$\times$4096 & 0.5 & Magnetic Field in photosphere \\
\hline
White Light Coronagraph & $\times$ & $\times$ & 400-850 nm & 2 - 30 & 1024$\times$1024 & 56 & Velocity of CMEs near the Sun \\
\hline
Heliospheric Imager & & $\times$ & 400-1000 nm & 130 - 216 & 1024$\times$1024 & 148 & Properties of the propagation of CMEs through interplanetary medium \\
\hline
EUV Imager & $\times$ & $\times$ & FeIX/X (17.4 nm) & $<$ 1.6 & 1024$\times$1024 & 3.2 & Coronal structures \\
\hline
\end{tabular}
 \end{table*}

 \begin{table*}
    \centering
    \caption[]{Constraints of the remote sensing instruments.}
    \label{Tab:opt2}
    \begin{tabular}{p{0.22\textwidth}p{0.04\textwidth}p{0.11\textwidth}p{0.05\textwidth}p{0.13\textwidth}p{0.2\textwidth}}
	\hline\hline
Instrument & Mass (kg) & Size (cm) & Power (W) & Data volume per day & Operation Temperature (K) \\
\hline
UV and IR coronagraph & 50 & 200$\times$60$\times$25 & 80 & \specialcell{4.7 GB (SCE)\\238 MB (SC80)} & \specialcell{223 (detector)\\303 (IR-polarimeter)} \\
Multi-Magnetic Imager & & 150$\times$70$\times$30 & 95 & 37.8 GB & \specialcell{233 (CCD)\\ 300 (tunable optics} \\
Magnetic Imager & 73 & 120$\times$85$\times$30 & 95 & 645 MB & \specialcell{233 (CCD)\\ 300 (tunable optics} \\
C2 Coronagraph & 15 & 140$\times$40$\times$32 & 5 & 3.8 MB & 193 \\
Heliospheric Imager & 15 & 65$\times$33$\times$20 & 10 & 0.4 MB & 193 \\
EUV Imager & 11 & 56$\times$15$\times$12.5 & 5 & 21.6 MB & 233 - 333 \\
\hline
\end{tabular}
\end{table*}

\subsection{Multi-channel Magnetic Imager (MMI)}
\label{sec:MMI}

Measurements of the vector magnetic field (Stokes Vector) of the photosphere and the chromosphere are performed by imaging different spectral lines in the visible spectrum. To measure the fields in the photosphere we use the iron lines at 630.15 nm and 603.25 nm. For the chromosphere the sodium lines D1 (589.592 nm) and D2 (588.995 nm) are used. For each one of the four lines, Full-Disk images are recorded by a Multichannel Magnetic Imager (MMI).

\begin{figure*}
\begin{center}
\includegraphics[width=0.8\textwidth]{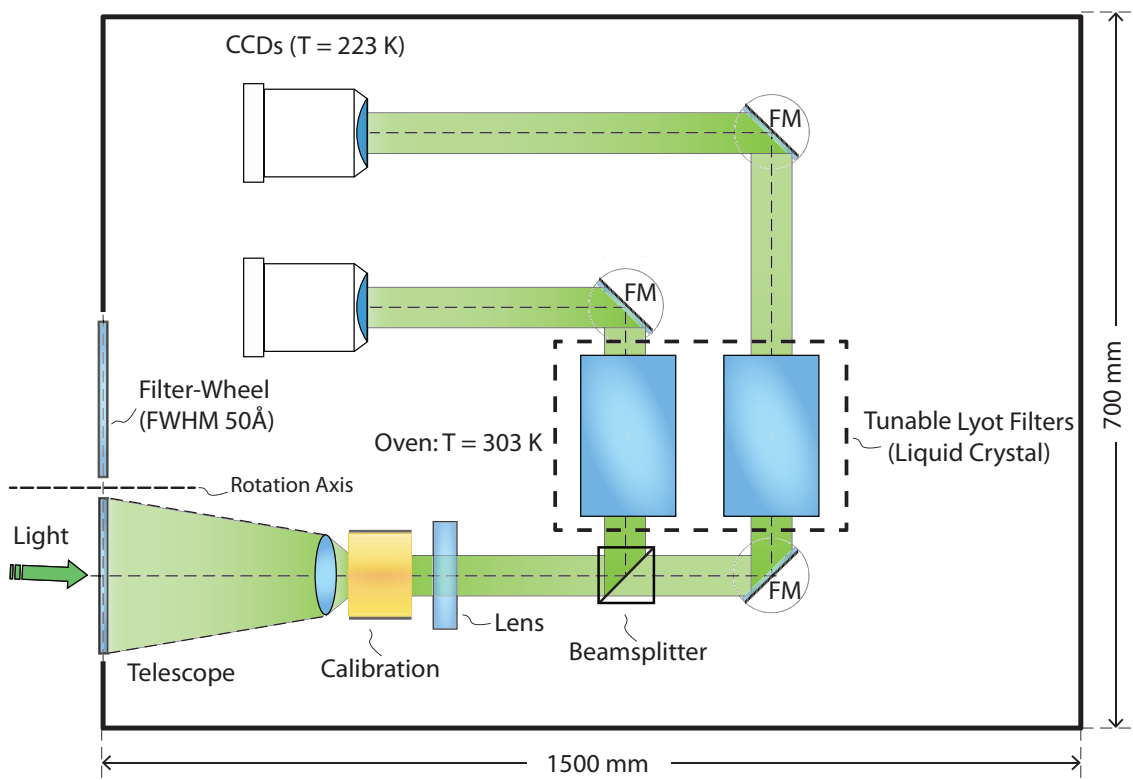}
\end{center}
\caption{Design of the new Multi-Magnetic Imager (MMI) instrument.}
\label{MMI}
\end{figure*}

Figure~\ref{MMI} shows the design of the new instrument. It is roughly based on existing imagers such as HMI (SDO) and ASPIICS (Proba 3). The refracting telescope with a diameter of 14 cm ensures the required spatial resolution. The two lines are measured in parallel. The selection of iron and sodium lines and the protection against overheating due to sun light is done by external filters (FWHM 5 nm) in a filter wheel. The different lines are selected using tunable liquid crystal Lyot filters which have to be heated to 300 $\pm$ 0.1 K.

Two CCD cameras (4096$\times$4096 pixel) with a resolution of 0.5 arcsec/pixel allow the ability to distinguish between different important zones within an active region. These are passively cooled to 233 K and take a series of 24 pictures for two spectral lines simultaneously within 2 min, before the filter wheel switches to the alternative lines. Each of these series is then preprocessed on-board to obtain images of continuum intensity, line of sight velocity and magnetic field components. A complete set of observations is obtained every four minutes.

\subsection{Photospheric Magnetic Imager (MI)}

The mission can monitor active regions in the sun for two weeks before they are aligned with the Earth. The Stokes vector of the photosphere is measured using the Fe I line at 617.3 nm. This type of observation has been already performed in previous missions. The MI is inspired by the HMI on SDO. A set of 24 images every hour allow to monitoring of the magnetic field components, line of sight velocity and continuum intensity of the photosphere. Details of the instrument characteristics are given in Table~\ref{Tab:opt}.

\subsection{Extreme Ultraviolet Imager (EUVI)}

Compared to the other parts of the solar atmosphere the corona emits light in fewer spectral lines. Current observations (STEREO, SDO, PROBA2) concentrate mainly on the Fe IX/X line in Extreme UV (EUV). In particular, coronal loops, which are indicators of closed magnetic field lines, are visible in this line.

In order to investigate the onset of flares and CMEs images of the whole solar disc in this EUV line are essential. Measurements of the intensity in this line are achieved with imaging devices that use a narrow-band filter. We use the SWAP EUV imager \citep[see][]{Seaton:2013} from the PROBA2 mission \citep[see][]{PROBA2} which takes images at 17.4 nm. Due to its novel and compact Ritchey-Chr\'etien scheme with an aperture of 33 mm it can be built in a compact form. It uses a new CCD technology and has a low power consumption of 5W. The time cadence can be controlled and is currently set between 110 s and 120 s in the PROBA2 mission.

The FOV ranges up to 1.6 R$_{s}$ measured from the disc center. We do not need to make additional adaptations of the SWAP instrument for our mission because it exactly fulfills the requirements. This allows observation of coronal dynamics on short time scales, e.g. at the onset of a CME.

\subsection{White Light Coronagraph}

White-light coronagraphs on SCE and SC80 are essential for this mission. They provide a view on the solar corona by blocking the much brighter light of the photosphere. This is achieved by an occulter disc.

The most important criterion for the quality of a coronagraph is its ability to suppress stray light from the lower solar atmosphere. Due to the sharp decrease in intensity of the outer corona the white light coronagraph has to cover an intensity range of several orders of magnitude.

We decided to use the heritage of the LASCO C3 coronagraph of the SOHO mission \citep[see][]{Brueckner:1995} in order to gain a FOV of 2 R$_{s}$ to 30 R$_{s}$ and cover a brightness range from 10$^{-8}$ to 10$^{-11}$ of the Mean Solar Brightness (MSB). The Cor2 Coronagraph of STEREO uses the same technology but has a smaller FOV. The LASCO C3 has now been in space for 18 years and still produces high-quality images of the outer corona and interplanetary space. The external occulter guarantees a low stray light contamination.

Due to the low density and brightness, higher exposure times are needed in comparison with other PAC2MAN mission imagers. The exposure time of LASCO C3 (about 19 seconds) and its time cadence (24 min) is enough to fulfill the requirements of the PAC2MAN mission. With the White Light Coronagraph we are able to observe the outer corona and the beginning of CME propagation, including their speed and their direction of propagation.

Limitations in the telemetry from SC80 will require the development of on-board processing algorithms that detect, analyze and transmit only the sections of the image that contain the Earth-directed CMEs. Such improvements should be part of subsequent mission design phases.

\subsection{Heliospheric Imager}

This instrument is present only in the SC80. Its design is based in the two Heliospheric Imagers installed on the STEREO spacecraft. The HI is designed for a stray light suppression of 10$^{-14}$ B0 (B0 = solar disc intensity) and a wide angle field of view. For this mission the HI-1 imaging system is removed completely and the HI-2 field of view is reduced to 42\degres \ with a pointing to 29\degres \ elongation. The other main characteristics of the HI are listed in Table~\ref{Tab:opt}.

 \subsection{In-Situ Instruments}
\label{sec:plasma}

Instruments are placed on the SCE spacecraft to measure the interplanetary plasma conditions near the Earth. The prediction of solar wind conditions and IMF orientation will help to define a set of improved input parameters for magnetospheric and ionospheric models. Due to telemetry constraints on the SC80 only one in-situ instrument, measuring the low energy solar wind, is included.

For these measurements we have three main instruments: The Low Energy Solar Wind Sensors (LEWiS), the High Energy Particle Sensors (HEPS), and the Fluxgate Magnetometer (MAG). These instruments are directly inherited from the Solar Orbiter particles package \citep[see][]{Mueller2013}. They cover a large array of energies, detect different types of particles, are lightweight and have a low power consumption.

Solar wind and high energy plasma characteristics can be accurately correlated with observations made by the optical instruments. Numerical models based on both measurements will improve our understanding of the effects of solar events on the near-Earth environment.

\subsubsection{Low Energy Solar Wind Sensors (LEWiS).} The first pack of 
instruments, LEWiS measures solar wind bulk properties leading to the detection of CMEs and interplanetary shocks. Solar wind properties (velocity, density, temperature, magnetic field) fluctuate constantly causing geomagnetic storms of different intensities on the Earth. The pack includes the three following instruments:

\begin{itemize}
\item {\bf The Electron Analyzer System.}

The Electron Analyzer System (EAS) measures the velocity distribution function of electrons. Electrons arrive to the instrument from any direction due to their large thermal spread. To provide a full 4$\pi$ sr field of view, the two sensors used have a $\pm$45\degres aperture deflection and are mounted orthogonal to each other in the shadow of the spacecraft at the end of an instrument boom. Each of the orthogonal sensors consists of a pair of top-hat electrostatic analyzers with aperture deflection plates.

\item {\bf The Proton-Alpha Sensor.}

The Proton-Alpha Sensor (PAS) measures the 3D velocity distribution function of the dominant solar wind species (protons and $\alpha$-particles). It consists of an electrostatic analyzer with an ion steering (IS) system at the aperture. It has an angular resolution $\leq2$\degres across a field of view of -17.5\degres to +47.5\degres by $\pm$ 22.5\degres about the solar direction.

This sensor is the only in-situ instrument installed in both spacecraft.

\item {\bf The Heavy Ion Sensor.}

The Heavy Ion Sensor (HIS) analyses the minor ion components of the solar wind, like major charge states of oxygen and iron, and 3D velocity distribution of some weakly ionized species (C$^{+}$, N$^{+}$,Mg$^{+}$, Si$^{+}$, Ne$^{+}$, He$^{+}$ etc). HIS measures five key properties: mass (range: 2-56 amu/q), charge, energy (0.5-100 keV/q for azimuth and 0.5-16 keV/q for elevation) and direction of incidence ($\theta, \varphi$).

To measure these properties the LEWiS/HIS is made up of four contiguous parts: the solar faced entrance aperture and ion steering (IS) directs incoming ions from different elevations/azimuths towards the electrostatic module, followed by a time-of-flight (TOF) telescope and solid state detectors (SSD).

In total it measures different energy ranges from a few eV to $\sim$5 keV for electrons and $\sim$0.2-100 keV/q for heavy ions and protons. It has a total mass of $\sim$15.9 kg. The LEWiS is inspired by similar instruments onboard Solar Orbiter \citep[see][]{Mueller2013}, STEREO \citep[see][]{Kaiser2008}, Ulysses \citep[see][]{Wenzel1992}, 
and ACE \citep[see][]{Bernath2005}.
\end{itemize}

\begin{table*}[h]
      \caption[]{Energy Range in keV or keV/nuc for the Sensors of the Low 
Energy Solar Wind Analyzer (LEWiS) for different energetic Particles.}
         \label{Tab:LEWiS}
    \begin{center}
    \begin{tabular}{cccc}
	\hline\hline
Sensor & Electrons & Protons & Heavy Ions (C$^{+}$, N$^{+}$,Mg$^{+}$, Si$^{+}$, 
Ne$^{+}$, He$^{+}$, etc.)\\
	\hline
EAS  & 0.001 -- 5  &  &  \\
PAS  &   &  0.2 -- 20 &   \\
HIS  &   &   & 0.5-100 for azimuth and  0.5-16  for elevation \\
\hline
\end{tabular}
\end{center}
\end{table*}

\subsubsection{High Energy Particle Sensor}
The High Energy Particle Sensor (HEPS) measures high energy particles caused by solar energetic events \citep[see][]{Miralles2011}. It  measures the full composition of the energetic particles, velocity dispersion and proton/electron intensities in various energy ranges from a few keV up to 200~MeV (see Table~\ref{Tab:HEPS})\citep[see][]{Mueller2013}. It combines five separate sensors which share a common data processing unit. The different sensors have different field-of-views and are described next:

\begin{itemize}
\item {\bf Supra-Thermal Electrons, Ions and Neutrals}

The first sensor of HEPS, the Supra-Thermal Electrons, Ions and Neutrals Telescope (STEIN), is a double-ended telescope. It uses a passively cooled silicon semiconductor detector (SSD) to measure supra-thermal particles with an energy range from 3--100~keV. To separate ions and electrons STEIN uses an electrostatic deflection system. The ions are stopped by the system up to an energy of 40~keV and neutrals up to 10~keV. The STEIN telescope has to be mounted on a boom and covers a field-of-view of 70$^{\circ}\times$60$^{\circ}$. The sensor is pointing into the direction of the Parker spiral.

\item {\bf Supra-thermal Ion Spectrograph}

The SIS detects heavy ions with an energy range of 8~keV/nuc--10~MeV/nuc, as well as ultra heavy ions in $^{3}$He rich solar flares below 1~MeV/nuc. It uses time-of-flight mass spectrometry \citep[see][]{Mueller2013}. The two telescopes with a FOV of 22$^{\circ}$ are pointing in the sun-ward and anti-sun-ward direction, detecting the particles when they pass through the entrance foil. The SSD detectors have a large detection area of $\approx$~12~cm$^{2}$.

\item {\bf High Energy Telescope}

The High Energy Telescope (HET) detects high-energy particles with energy ranges of 10--100~MeV for protons, 20--200~MeV/nuc for heavier ions and 300~keV--20~MeV for electrons. It deals with the energy ranges of larger solar events and allows a separation of $^{3}$He/$^{4}$He isotope ratio of about 1$\%$.  The sensor uses a combination of solid-state detectors and a scintillator calorimeter. This allows us to use the dE/dx vs. total E technique for particle identification and energy measurements \citep[see][]{Mueller2013}. Two double-ended sensors are used, one pointing sun-ward and the other anti-sun-ward. The HET has a FOV of 50$^{\circ}$. 

\item {\bf Low Energy Telescope}

The Low Energy Telescope (LET) measures heavy ions from H to Ni with an energy range of 1.5--60~MeV/nuc. With the LET it is possible to resolve $^{3}$He and multiple heavy ion species (Ne and Mg) in order to identify particle sources. The instrument uses six small stacks of silicon detectors. To get 3D information the sensor has three telescopes with an angular separation of 60$^{\circ}$. The LET closes the energy gap between SIS and HET for heavy ion measurements. 

\item {\bf Electron Proton Telescope}

The Electron Proton Telescope (EPT) measures electrons (energy range: 20~keV--700~keV), protons (energy range: 20~keV--9~MeV) and their anisotropies with the magnet/foil-technique \citep[see][]{Mueller2013}. One of the two double-ended telescopes points along the Parker spiral and the other points 45$^{\circ}$ out of the ecliptic. The instrument closes the energy gap of electrons between STEIN and HET and the energy gap between STEIN and LET for proton energy ranges. 
\end{itemize}

\begin{table*}
      \caption[]{Energy Range in MeV or MeV/nuc for the Sensors of the High 
Energy Particle Sensor (HEPS) for different energetic Particles 
\citep[see][]{Mueller2013}.}
         \label{Tab:HEPS}
    \begin{center}
    \begin{tabular}{cccccccc}
	\hline\hline
Sensor & Electrons & Protons & He &  $^{3}$He & CNO & NeMgSi & Fe\\
	\hline
STEIN & 0.002 -- 0.1  & 0.003 -- 0.1 & -- & -- & -- & -- & --\\
EPT   & 0.02 -- 0.4 & 0.02 -- 7  & -- & -- & -- & -- & -- \\
SIS   & -- & -- & 0.02 -- 8 & 0.02 -- 9 & 0.01 -- 10 & 0.01 -- 10  & 0.01 -- 9 
\\
LET   & -- & 1.5 -- 20 & 1.4 -- 19 & 1.8 -- 21 & 2.5 -- 40 & 3 -- 50 & 3 -- 70 
\\
HET   & 0.3--20 & 10--100 & 10--100 & 12 -- 120 & 20 -- 200 & 30 -- 200 & 40 -- 
200 \\
\hline
\end{tabular}
\end{center}
 \end{table*}

In total HEPS measures different energy ranges from few keV to 200~MeV for heavy ions, protons, and electrons. It has a total mass of 16.1~kg and an operation temperature of 233--333~K. The HEPS is inspired by Solar Orbiter \citep[see][]{Mueller2013}, STEREO \citep[see][]{Kaiser2008}, SOHO \citep[see][]{Domingo1995}, and ACE \citep[see][]{Bernath2005}.

\subsubsection{Magnetometer} 
The Magnetometer (MAG) measures the heliospheric magnetic field at the L1 point with a measuring range of $\pm$ 500 nT. Using a dual fluxgate sensor the instrument reaches high precision of $\approx$~4~pT \citep[see][]{Mueller2013}. Two detectors are used: one in the vicinity of the spacecraft and the second far away mounted in a boom. Interference from the spacecraft is extracted from the heliospheric magnetic field using an advanced dual sensor technique: it has been observed that beyond a certain distance, the induced magnetic field of the spacecraft can be assumed dipolar. This field can be separated from the background field by placing magnetometers at two different radial locations aligned with the center of the dipole \citep[see][]{Ness:1971,Acuna:2002}. Both detectors are mounted on the instrument boom in the shadow of the spacecraft. MAG has high time-resolution of 100 vectors/s.

\section{Conclusions}

The space mission presented in this paper is the result of the work performed by 15 young scientists and engineers in the framework of the ESA Alpbach Summer School 2013. All the technical and scientific solutions were developed in a time frame of 10 days.

There is currently a need for more precise measurements of the energetic content of active regions of the Sun. We propose in section \ref{sec:scireq} that a continuous surveillance of the vector magnetic field at different altitudes in the solar atmosphere can be used to improve the modeling of the free magnetic energy that triggers CMEs and flares.

The objective of the mission is to identify several days in advance when eruptive events will take place on the Sun and then forecast their Earth-arrival once they erupt. Full coverage of the space weather event can be obtained by additional in-situ measurements of the plasma characteristics of the solar wind near the Earth.

To attain the objectives presented in section \ref{sec:objectives}, we place two spacecraft in different orbits: Spacecraft Earth (SCE) is located in an orbit around the Lagrangian L1 point, and Spacecraft 80 (SC80) is located in a heliosynchronous orbit at 1 AU trailing the Earth by 80\degres. Measurements of the vector magnetic field at different altitudes (photosphere, chromosphere and corona) are performed by a series of optical instruments. Two of the instruments where designed from scratch and are presented in sections \ref{sec:UVIRC} and \ref{sec:MMI}. The choice of spectral lines presented in section \ref{sec:lines} allows the detection of magnetic field components by interpreting the Hanle effect, never used before in a spaceborne instrument. The spectroscopic view obtained from the two spacecraft allows the instrument to perform accurate combined observations of the three dimensional magnetic structures in the solar disk and in the off-limb region.

Plasma instruments mounted in the SCE measure the properties of the solar wind ahead of the Earth. Correlations between observations in the Sun and plasma characteristics will allow building new models for the forecasting of geoeffectiveness of space weather events. Section \ref{sec:plasma} shows details on the types of instruments used to detect electrons, protons and heavy ions at different energy ranges. Fluxgate magnetometers measure the three components of the interplanetary magnetic field.

The mission profile and details about the engineering solutions proposed are presented in section \ref{sec:mission}. We payed special attention in designing the spacecraft to include accurate constraints in terms of mass, control methods, communications, power supply, thermal balance and cost.

For a total estimated cost of 1180 M euros we have designed a space weather mission that will track continuously, for 6 years, the evolution of eruptive events that affect human life and technology. CMEs and flares will be followed from days before their formation on the Sun to their effects in the Earth environment. This mission is a significant step forward for the forecasting of eruptive solar events and their geoeffectiveness.

\section{Acknowledgements}

The publication of the work presented in this paper received funding from the European Commission's Seventh Framework Program (FP7/2007-2013) under the grant agreement eHeroes (project 284461, eheroes.eu).

The main work for this project was performed during the ESA Summer School Alpbach 2013. We would like to thank the organizers of the summer school, the European Space Agency (ESA), the Austrian Research Promotion Agency (FFG), the International Space Science Institute (ISSI), the Association of Austrian Space Industries (AUSTROSPACE), and the Summer School director Michaela Gitsch.

Each participant would like to acknowledge the financial support for the participation in the Alpbach Summer School 2013 to the following institutions: CNES, the LABEX ESP and ANR for their support through the ``investissement d'avenir'' program (Sophie Musset), DLR (Ren\'e Kiefer), the Swedish National Space Board (Viktor Andersson) and the Swiss Committee on Space Research, Swiss Academy of Sciences (Mohammad Madi).

We wish to thank our summer school tutors, Marcus Hallmann and Manuela Temmer for their excellent supervision and support. In addition we would like to thank Peter Falkner, Margit Haberreiter, Andre Balogh, Christian Erd, Juha-Pekka Luntama, Anik de Groof, Volker Bothmer, Denis Moura, G\"unther Reitz, Roger Bonnet and all the tutors for the help they provided during the summer school, their feedback and good ideas.

The editor thanks Bernard Jackson and an anonymous referee for their assistance in evaluating this paper.

\bibliography{swsc140009-biblio}

\begin{thebibliography}{50}
\providecommand{\natexlab}[1]{#1}
\providecommand{\url}[1]{\texttt{#1}}
\providecommand{\urlprefix}{}

\bibitem[{Acuna(2002)}]{Acuna:2002}
Acuna, M.H.
\newblock Space-based magnetometers.
\newblock \emph{Review of Scientific Instruments}, \textbf{73~(11)},
  3717--3736, 2002.
\newblock
  \urlprefix\url{http://scitation.aip.org/content/aip/journal/rsi/73/11/10.1063/1.1510570}.

\bibitem[{{Alissandrakis} and {Chiuderi Drago}(1995)}]{Alissandrakis:1995}
{Alissandrakis}, C.A., and {Chiuderi Drago}, F.
\newblock {Coronal Magnetic Fields from Faraday Rotation Observations}.
\newblock \emph{\solphys}, \textbf{160}, 171--179, 1995.

\bibitem[{{Amari} et~al.(1997){Amari}, {Aly}, {Luciani}, {Boulmezaoud}, and
  {Mikic}}]{Amari:1997}
{Amari}, T., {Aly}, J.J., {Luciani}, J.F., {Boulmezaoud}, T.Z., and {Mikic}, Z.
\newblock {Reconstructing the Solar Coronal Magnetic Field as a Force-Free
  Magnetic Field}.
\newblock \emph{\solphys}, \textbf{174}, 129--149, 1997.

\bibitem[{{Aulanier} et~al.(2013)}]{Aulanier:2013}
{Aulanier}, G., et~al.
\newblock {The standard flare model in three dimensions. II. Upper limit on
  solar flare energy}.
\newblock \emph{\aap}, \textbf{549}, A66, 2013.

\bibitem[{{Bernath} et~al.(2005)}]{Bernath2005}
{Bernath}, P.F., et~al.
\newblock {Atmospheric Chemistry Experiment (ACE): Mission overview}.
\newblock \emph{\grl}, \textbf{32}, L15S01, 2005.

\bibitem[{{Bommier} and {Gelly}(2014)}]{Bommier:2014}
{Bommier}, V., and {Gelly}, B.
\newblock {Magnetometry from HINODE/SOT/SP data: solving the fundamental
  ambiguity from the Fe I 6301/6302 line pair inversion}.
\newblock \emph{\pasj}, 2014.

\bibitem[{{Bommier} and {Sahal-Brechot}(1982)}]{Bommier:1982}
{Bommier}, V., and {Sahal-Brechot}, S.
\newblock {The Hanle effect of the coronal L-alpha line of hydrogen -
  Theoretical investigation}.
\newblock \emph{\solphys}, \textbf{78}, 157--178, 1982.

\bibitem[{{Brueckner} et~al.(1995)}]{Brueckner:1995}
{Brueckner}, G.E., et~al.
\newblock {The Large Angle Spectroscopic Coronagraph (LASCO)}.
\newblock \emph{\solphys}, \textbf{162}, 357--402, 1995.

\bibitem[{Cane and Richardson(2003)}]{Cane:2003}
Cane, H.V., and Richardson, I.G.
\newblock Interplanetary coronal mass ejections in the near-Earth solar wind
  during 1996--2002.
\newblock \emph{Journal of Geophysical Research: Space Physics},
  \textbf{108~(A4)}, n/a--n/a, 2003.
\newblock ISSN 2156-2202.
\newblock \urlprefix\url{http://dx.doi.org/10.1029/2002JA009817}.

\bibitem[{{Cheung} and {DeRosa}(2012)}]{Cheung:2012}
{Cheung}, M.C.M., and {DeRosa}, M.L.
\newblock {A Method for Data-driven Simulations of Evolving Solar Active
  Regions}.
\newblock \emph{\apj}, \textbf{757}, 147, 2012.

\bibitem[{{Chevalier} and {Lambert}(1969)}]{Chevalier:1969}
{Chevalier}, R.A., and {Lambert}, D.L.
\newblock {The Excitation of the Forbidden Coronal Lines. I: Fe XIII
  {$\lambda$}{$\lambda$} 10747, 10798 and 3388}.
\newblock \emph{\solphys}, \textbf{10}, 115--134, 1969.

\bibitem[{{Cliver} and {Dietrich}(2013)}]{Cliver:2013}
{Cliver}, E.W., and {Dietrich}, W.F.
\newblock {The 1859 space weather event revisited: limits of extreme activity}.
\newblock \emph{Journal of Space Weather and Space Climate}, \textbf{3~(26)},
  A31, 2013.

\bibitem[{Curto and Gaya-Piqu\'e(2009)}]{Curto:2009}
Curto, J.J., and Gaya-Piqu\'e, L.R.
\newblock Geoeffectiveness of solar flares in magnetic crochet (sfe)
  production: I{\^a}€''Dependence on their spectral nature and position on
  the solar disk.
\newblock \emph{Journal of Atmospheric and Solar-Terrestrial Physics},
  \textbf{71}, 1695 -- 1704, 2009.
\newblock ISSN 1364-6826.
\newblock
  \urlprefix\url{http://www.sciencedirect.com/science/article/pii/S1364682609001205}.

\bibitem[{{Demoulin} et~al.(1997){Demoulin}, {Henoux}, {Mandrini}, and
  {Priest}}]{Demoulin:1997}
{Demoulin}, P., {Henoux}, J.C., {Mandrini}, C.H., and {Priest}, E.R.
\newblock {Can we Extrapolate a Magnetic Field when its Topology is Complex?}
\newblock \emph{\solphys}, \textbf{174}, 73--89, 1997.

\bibitem[{{Domingo} et~al.(1995){Domingo}, {Fleck}, and {Poland}}]{Domingo1995}
{Domingo}, V., {Fleck}, B., and {Poland}, A.I.
\newblock {The SOHO Mission: an Overview}.
\newblock \emph{\solphys}, \textbf{162}, 1--37, 1995.

\bibitem[{{Emslie} et~al.(2004)}]{Emslie:2004}
{Emslie}, A.G., et~al.
\newblock {Energy partition in two solar flare/CME events}.
\newblock \emph{Journal of Geophysical Research: Space Physics}, \textbf{109},
  A10104, 2004.

\bibitem[{Feynman and Gabriel(2000)}]{Feynman:2000}
Feynman, J., and Gabriel, S.B.
\newblock On space weather consequences and predictions.
\newblock \emph{Journal of Geophysical Research: Space Physics},
  \textbf{105~(A5)}, 10543--10564, 2000.
\newblock ISSN 2156-2202.
\newblock \urlprefix\url{http://dx.doi.org/10.1029/1999JA000141}.

\bibitem[{Feynman and Martin(1995)}]{Feynman:1995}
Feynman, J., and Martin, S.F.
\newblock The initiation of coronal mass ejections by newly emerging magnetic
  flux.
\newblock \emph{Journal of Geophysical Research: Space Physics},
  \textbf{100~(A3)}, 3355--3367, 1995.
\newblock ISSN 2156-2202.
\newblock \urlprefix\url{http://dx.doi.org/10.1029/94JA02591}.

\bibitem[{{Flower} and {Pineau des Forets}(1973)}]{Flower:1973}
{Flower}, D.R., and {Pineau des Forets}, G.
\newblock {Excitation of the Fe XIII Spectrum in the Solar Corona}.
\newblock \emph{\aap}, \textbf{24}, 181, 1973.

\bibitem[{{Gabriel}(1971)}]{Gabriel:1971}
{Gabriel}, A.H.
\newblock {Measurements on the Lyman Alpha Corona (Papers presented at the
  Proceedings of the International Symposium on the 1970 Solar Eclipse, held in
  Seattle, U. S. A. , 18-21 June, 1971.)}.
\newblock \emph{\solphys}, \textbf{21}, 392--400, 1971.

\bibitem[{{Gary} and {Hurford}(1994)}]{Gary:1994}
{Gary}, D.E., and {Hurford}, G.J.
\newblock {Coronal temperature, density, and magnetic field maps of a solar
  active region using the Owens Valley Solar Array}.
\newblock \emph{\apj}, \textbf{420}, 903--912, 1994.

\bibitem[{{Gopalswamy} et~al.(2003){Gopalswamy}, {Lara}, {Yashiro}, {Nunes},
  and {Howard}}]{Gopalswamy:2003}
{Gopalswamy}, N., {Lara}, A., {Yashiro}, S., {Nunes}, S., and {Howard}, R.A.
\newblock {Coronal mass ejection activity during solar cycle 23}.
\newblock In A.~{Wilson}, editor, \emph{Solar Variability as an Input to the
  Earth's Environment}, 2003, volume 535 of \emph{ESA Special Publication},
  403--414.

\bibitem[{Gopalswamy et~al.(2007)Gopalswamy, Yashiro, and
  Akiyama}]{Gopalswamy:2007}
Gopalswamy, N., Yashiro, S., and Akiyama, S.
\newblock Geoeffectiveness of halo coronal mass ejections.
\newblock \emph{Journal of Geophysical Research: Space Physics}, \textbf{112},
  2007.
\newblock ISSN 2156-2202.
\newblock \urlprefix\url{http://dx.doi.org/10.1029/2006JA012149}.

\bibitem[{{Harder} et~al.(2005){Harder}, {Lawrence}, {Fontenla}, {Rottman}, and
  {Woods}}]{Harder:2005}
{Harder}, J., {Lawrence}, G., {Fontenla}, J., {Rottman}, G., and {Woods}, T.
\newblock {The Spectral Irradiance Monitor: Scientific Requirements, Instrument
  Design, and Operation Modes}.
\newblock \emph{\solphys}, \textbf{230}, 141--167, 2005.

\bibitem[{{Judge} et~al.(2001){Judge}, {Casini}, {Tomczyk}, {Edwards}, and
  {Francis}}]{Judge:2001}
{Judge}, P.G., {Casini}, R., {Tomczyk}, S., {Edwards}, D.P., and {Francis}, E.
\newblock {Coronal Magnetometry: A Feasibility Study}.
\newblock \emph{NASA STI/Recon Technical Report N}, \textbf{2}, 27999, 2001.

\bibitem[{{Kaiser} et~al.(2008)}]{Kaiser2008}
{Kaiser}, M.L., et~al.
\newblock {The STEREO Mission: An Introduction}.
\newblock \emph{\ssr}, \textbf{136}, 5--16, 2008.

\bibitem[{{Kretzschmar}(2011)}]{kret:2011}
{Kretzschmar}, M.
\newblock {The Sun as a star: observations of white-light flares}.
\newblock \emph{\aap}, \textbf{530}, A84, 2011.

\bibitem[{{Lin} et~al.(2004){Lin}, {Kuhn}, and {Coulter}}]{Lin:2004}
{Lin}, H., {Kuhn}, J.R., and {Coulter}, R.
\newblock {Coronal Magnetic Field Measurements}.
\newblock \emph{\apjl}, \textbf{613}, L177--L180, 2004.

\bibitem[{Luhmann et~al.(1998)Luhmann, Gosling, Hoeksema, and
  Zhao}]{Luhmann:1998}
Luhmann, J.G., Gosling, J.T., Hoeksema, J.T., and Zhao, X.
\newblock The relationship between large-scale solar magnetic field evolution
  and coronal mass ejections.
\newblock \emph{Journal of Geophysical Research: Space Physics},
  \textbf{103~(A4)}, 6585--6593, 1998.
\newblock ISSN 2156-2202.
\newblock \urlprefix\url{http://dx.doi.org/10.1029/97JA03727}.

\bibitem[{Mancuso and Spangler(1999)}]{Mancuso:1999}
Mancuso, S., and Spangler, S.R.
\newblock Coronal Faraday Rotation Observations: Measurements and Limits on
  Plasma Inhomogeneities.
\newblock \emph{The Astrophysical Journal}, \textbf{525~(1)}, 195, 1999.
\newblock \urlprefix\url{http://stacks.iop.org/0004-637X/525/i=1/a=195}.

\bibitem[{{McClymont} et~al.(1997){McClymont}, {Jiao}, and
  {Mikic}}]{McClymont:1997}
{McClymont}, A.N., {Jiao}, L., and {Mikic}, Z.
\newblock {Problems and Progress in Computing Three-Dimensional Coronal Active
  Region Magnetic Fields from Boundary Data}.
\newblock \emph{\solphys}, \textbf{174}, 191--218, 1997.

\bibitem[{{Millard} et~al.(2004){Millard}, {Lemaire}, and {Vial}}]{lyot}
{Millard}, A., {Lemaire}, P., and {Vial}, J.C.
\newblock {EUV imager and spectrometer for Lyot and Solar Orbiter space
  missions}.
\newblock In B.~{Warmbein}, editor, \emph{5th International Conference on Space
  Optics}, 2004, volume 554 of \emph{ESA Special Publication}, 351--354.

\bibitem[{Miralles and Almeida(2011)}]{Miralles2011}
Miralles, M., and Almeida, J.
\newblock \emph{The Sun, the Solar Wind, and the Heliosphere}.
\newblock IAGA Special Sopron Book Series. Springer, 2011.
\newblock ISBN 9789048197873.

\bibitem[{{M{\"u}ller} et~al.(2013){M{\"u}ller}, {Marsden}, {St.~Cyr}, and
  {Gilbert}}]{Mueller2013}
{M{\"u}ller}, D., {Marsden}, R.G., {St.~Cyr}, O.C., and {Gilbert}, H.R.
\newblock Solar Orbiter.
\newblock \emph{Solar Physics}, \textbf{285~(1-2)}, 25--70, 2013.
\newblock ISSN 0038-0938.
\newblock \urlprefix\url{http://dx.doi.org/10.1007/s11207-012-0085-7}.

\bibitem[{Mustajab and Badruddin(2013)}]{Mustajab:2013}
Mustajab, F., and Badruddin.
\newblock Relative geo-effectiveness of coronal mass ejections with distinct
  features in interplanetary space.
\newblock \emph{Planetary and Space Science}, \textbf{82--83~(0)}, 43 -- 61,
  2013.
\newblock ISSN 0032-0633.
\newblock
  \urlprefix\url{http://www.sciencedirect.com/science/article/pii/S0032063313000706}.

\bibitem[{Ness et~al.(1971)Ness, Behannon, Lepping, and Schatten}]{Ness:1971}
Ness, N.F., Behannon, K.W., Lepping, R.P., and Schatten, K.H.
\newblock Use of two magnetometers for magnetic field measurements on a
  spacecraft.
\newblock \emph{Journal of Geophysical Research}, \textbf{76~(16)}, 3564--3573,
  1971.
\newblock ISSN 2156-2202.
\newblock \urlprefix\url{http://dx.doi.org/10.1029/JA076i016p03564}.

\bibitem[{Owens and Crooker(2006)}]{Owens:2006}
Owens, M.J., and Crooker, N.U.
\newblock Coronal mass ejections and magnetic flux buildup in the heliosphere.
\newblock \emph{Journal of Geophysical Research: Space Physics},
  \textbf{111~(A10)}, n/a--n/a, 2006.
\newblock ISSN 2156-2202.
\newblock \urlprefix\url{http://dx.doi.org/10.1029/2006JA011641}.

\bibitem[{{Patzold} et~al.(1987)}]{Patzold:1987}
{Patzold}, M., et~al.
\newblock {The mean coronal magnetic field determined from HELIOS Faraday
  rotation measurements}.
\newblock \emph{\solphys}, \textbf{109}, 91--105, 1987.

\bibitem[{{Peter} et~al.(2012)}]{SolmeX}
{Peter}, H., et~al.
\newblock {Solar magnetism eXplorer (SolmeX). Exploring the magnetic field in
  the upper atmosphere of our closest star}.
\newblock \emph{Experimental Astronomy}, \textbf{33}, 271--303, 2012.

\bibitem[{{Raouafi} et~al.(2009){Raouafi}, {Solanki}, and
  {Wiegelmann}}]{Raouafi:2009}
{Raouafi}, N.E., {Solanki}, S.K., and {Wiegelmann}, T.
\newblock {Hanle Effect Diagnostics of the Coronal Magnetic Field: A Test Using
  Realistic Magnetic Field Configurations}.
\newblock In S.V. {Berdyugina}, K.N. {Nagendra}, and R.~{Ramelli}, editors,
  \emph{Solar Polarization 5: In Honor of Jan Stenflo}, 2009, volume 405 of
  \emph{Astronomical Society of the Pacific Conference Series}, 429.

\bibitem[{Reinard et~al.(2010)Reinard, Henthorn, Komm, and Hill}]{Reinard:2010}
Reinard, A.A., Henthorn, J., Komm, R., and Hill, F.
\newblock Evidence That Temporal Changes in Solar Subsurface Helicity Precede
  Active Region Flaring.
\newblock \emph{The Astrophysical Journal Letters}, \textbf{710~(2)}, L121,
  2010.
\newblock \urlprefix\url{http://stacks.iop.org/2041-8205/710/i=2/a=L121}.

\bibitem[{{Sahal-Brechot}(1981)}]{Sahal:1981}
{Sahal-Brechot}, S.
\newblock {The Hanle effect applied to magnetic field diagnostics.}
\newblock \emph{\ssr}, \textbf{29}, 391--401, 1981.

\bibitem[{{Santandrea} et~al.(2013)}]{PROBA2}
{Santandrea}, S., et~al.
\newblock {PROBA2: Mission and Spacecraft Overview}.
\newblock \emph{\solphys}, \textbf{286}, 5--19, 2013.

\bibitem[{{Scherrer} et~al.(2012)}]{Scherrer:2012}
{Scherrer}, P.H., et~al.
\newblock {The Helioseismic and Magnetic Imager (HMI) Investigation for the
  Solar Dynamics Observatory (SDO)}.
\newblock \emph{\solphys}, \textbf{275}, 207--227, 2012.

\bibitem[{{Schrijver} et~al.(2012)}]{Schrijver:2012}
{Schrijver}, C.J., et~al.
\newblock {Estimating the frequency of extremely energetic solar events, based
  on solar, stellar, lunar, and terrestrial records}.
\newblock \emph{Journal of Geophysical Research: Space Physics}, \textbf{117},
  A08103, 2012.

\bibitem[{{Seaton} et~al.(2013)}]{Seaton:2013}
{Seaton}, D.B., et~al.
\newblock {The SWAP EUV Imaging Telescope Part I: Instrument Overview and
  Pre-Flight Testing}.
\newblock \emph{\solphys}, \textbf{286}, 43--65, 2013.

\bibitem[{{Sun} et~al.(2012)}]{Sun:2012}
{Sun}, X., et~al.
\newblock {Evolution of Magnetic Field and Energy in a Major Eruptive Active
  Region Based on SDO/HMI Observation}.
\newblock \emph{\apj}, \textbf{748}, 77, 2012.

\bibitem[{{Tomczyk} et~al.(2008)}]{Tomczyk:2008}
{Tomczyk}, S., et~al.
\newblock {An Instrument to Measure Coronal Emission Line Polarization}.
\newblock \emph{\solphys}, \textbf{247}, 411--428, 2008.

\bibitem[{{Wenzel} et~al.(1992){Wenzel}, {Marsden}, {Page}, and
  {Smith}}]{Wenzel1992}
{Wenzel}, K.P., {Marsden}, R.G., {Page}, D.E., and {Smith}, E.J.
\newblock {The ULYSSES Mission}.
\newblock \emph{\aaps}, \textbf{92}, 207, 1992.

\bibitem[{{Woods} et~al.(2006){Woods}, {Kopp}, and {Chamberlin}}]{Woods:2006}
{Woods}, T.N., {Kopp}, G., and {Chamberlin}, P.C.
\newblock {Contributions of the solar ultraviolet irradiance to the total solar
  irradiance during large flares}.
\newblock \emph{Journal of Geophysical Research: Space Physics}, \textbf{111},
  A10S14, 2006.

\end{thebibliography}
\bibliographystyle{jswsc}


\end{document}